\documentclass[11pt]{article}
\usepackage{moriond,epsfig}

\def\bea{\begin{eqnarray}}
\def\eea{\end{eqnarray}}
\def\beq{\begin{equation}}
\def\eeq{\end{equation}}

\newcommand{\ro}{\mbox{{\boldmath
$\rho$}}}

\newcommand{\pb}{\mbox{{\bf
p}}}

\newcommand{\bb}{{{\bf b}}}

\def\lsim{\mathrel{\rlap{\lower4pt\hbox{\hskip1pt$\sim$}}
    \raise1pt\hbox{$<$}}}         
\def\gsim{\mathrel{\rlap{\lower4pt\hbox{\hskip1pt$\sim$}}
    \raise1pt\hbox{$>$}}}         

\begin{document}
\vspace*{4cm}
\title{
JET QUENCHING FROM RHIC TO LHC 
}

\author{B.G. ZAKHAROV }

\address{L.D. Landau Institute for Theoretical Physics,
        GSP-1, 117940,\\ Kosygina Str. 2, 117334 Moscow, Russia}

\maketitle\abstracts{
We perform a joint analysis of the
data from PHENIX at RHIC and ALICE at LHC on the nuclear modification factor $R_{AA}$. 
The computations are performed within the light-cone path integral approach to induced gluon emission.
Our results show that slow variation of $R_{AA}$ from RHIC to LHC
energies indicates that the QCD coupling constant is suppressed
in the quark-gluon plasma produced at LHC.
}


\noindent{\bf 1.} 
One of the most striking results of experiments at RHIC 
is strong suppression of high-$p_{T}$ hadrons in
$AA$-collisions \cite{RHIC_data} 
(called ``jet quenching'').
Recently, a similar effect has been observed 
in the ALICE experiment at LHC \cite{ALICE} for
$Pb+Pb$ collisions at $\sqrt{s}=2.76$ TeV.
The most natural reason for this phenomenon is parton energy loss
(radiative and collisional) in the hot quark-gluon 
plasma (QGP) produced in the initial stage of $AA$-collisions.
It is of great interest to perform a joint analysis of the RHIC and LHC data.
It is interesting since variation of the nuclear modification factor 
$R_{AA}$ from RHIC to LHC energies should not be very sensitive to 
the systematic theoretical uncertainties that are rather large. 
These uncertainties come mostly from 
multiple induced gluon emission.
The available theoretical approaches to 
radiative induced gluon emission \cite{BDMPS,LCPI,BSZ,W1,GLV1,AMY}
are restricted to one gluon emission, and the multiple gluon emission
is usually evaluated in the approximation of independent gluon 
radiation \cite{BDMS_RAA}. 

In this talk, I will present results of an analysis of the 
data on $R_{AA}$ for $Au+Au$ collisions
at $\sqrt{s}=200$ GeV from PHENIX \cite{PHENIX08} and 
for $Pb+Pb$ collisions at $\sqrt{s}=2.76$ TeV from ALICE \cite{ALICE}.
The analysis is based on the
light-cone path integral (LCPI) approach \cite{LCPI}. We evaluate the nuclear 
modification factor using the method developed in \cite{raa08}.
A major purpose of this analysis is to decide whether the variation
of $R_{AA}$ from RHIC to LHC indicates that the QCD coupling constant
becomes smaller in the plasma produced at LHC, which is hotter
than that at RHIC.

\vspace{.2cm}
\noindent{\bf 2.}
The nuclear modification factor $R_{AA}$ for a given impact parameter $b$
can be written as
\beq
R_{AA}(b)=\frac{{dN(A+A\rightarrow h+X)}/{d\pb_{T}dy}}
{T_{AA}(b){d\sigma(N+N\rightarrow h+X)}/{d\pb_{T}dy}}\,.
\label{eq:10}
\eeq
Here $\pb_{T}$ is the hadron transverse momentum, $y$ is rapidity (we
consider the central region $y=0$), 
$T_{AA}(b)=\int d\ro T_{A}(\ro) T_{A}(\ro-\bb)$, $T_{A}$ is the nucleus 
profile function. The differential yield for high-$p_{T}$ hadron production in 
$AA$-collision can be written in the form 
\beq
\frac{dN(A+A\rightarrow h+X)}{d\pb_{T} dy}=\int d\ro T_{A}(\ro)T_{A}(\ro-\bb)
\frac{d\sigma_{m}(N+N\rightarrow h+X)}{d\pb_{T} dy}\,,
\label{eq:20}
\eeq
where ${d\sigma_{m}(N+N\rightarrow h+X)}/{d\pb_{T} dy}$ is the medium-modified
cross section for the $N+N\rightarrow h+X$ process.
Similarly to the ordinary pQCD formula, we write it as
\beq
\frac{d\sigma_{m}(N+N\rightarrow h+X)}{d\pb_{T} dy}=
\sum_{i}\int_{0}^{1} \frac{dz}{z^{2}}
D_{h/i}^{m}(z, Q)
\frac{d\sigma(N+N\rightarrow i+X)}{d\pb_{T}^{i} dy}\,.
\label{eq:30}
\eeq
Here $\pb_{T}^{i}=\pb_{T}/z$ is the parton transverse momentum, 
${d\sigma(N+N\rightarrow i+X)}/{d\pb_{T}^{i} dy}$ is the
hard cross section,
$D_{h/i}^{m}$ is the medium-modified fragmentation function (FF)
for transition of a parton $i$ into the observed hadron $h$.
For the parton virtuality scale $Q$ we take the parton transverse
momentum $p^{i}_{T}$.
We assume that hadronization occurs outside of the QGP.
For jets with $E\lsim 100$ GeV the hadronization 
scale, $\mu_h$, is relatively small.
Indeed, 
one can easily show that the $L$ dependence of the parton virtuality
reads $Q^{2}(L)\sim \max{(Q/L,Q_0^{2})}$, where  
$Q_{0}\sim 1-2$ GeV is some minimal nonperturbative scale.
For RHIC and LHC, 
when $\tau_{QGP}\sim R_A$ ($\tau_{QGP}$ is the typical lifetime/size of the QGP, 
$R_A$ is the nucleus radius),
it gives $\mu_{h}\sim Q_{0}$ (for  
$E\lsim 100$ GeV).
Then we can write
\beq
D_{h/i}^{m}(z,Q)\approx\int_{z}^{1} \frac{dz'}{z'}D_{h/j}(z/z',Q_{0})
D_{j/i}^{m}(z',Q_{0},Q)\,,
\label{eq:40}
\eeq
where 
$D_{h/j}(z,Q_{0})$ is the vacuum FF, and
$D_{j/i}^{m}(z',Q_{0},Q)$ is the medium-modified FF
for transition of the initial parton $i$ with virtuality $Q$
to a parton $j$ with virtuality $Q_{0}$.
For partons with $E\lsim 100$ GeV the typical length scale dominating 
the energy loss in the DGLAP stage
is relatively small $\sim 0.3-1$ fm \cite{raa08}.
This length is of the order of the formation time of the QGP
$\tau_{0}\sim 0.5$ fm. Since the induced radiation stage
occurs at larger length scale $l\sim \tau_{0}\div \tau_{QGP}$, 
to the first approximation
one can ignore the overlap of the DGLAP and induced radiation stages
at all \cite{raa08}. Then we can write
\beq
D_{j/i}^{m}(z,Q_{0},Q)=\int_{z}^{1} \frac{dz'}{z'}D_{j/l}^{ind}(z/z',E_{l})
D_{l/i}^{DGLAP}(z',Q_{0},Q)\,,
\label{eq:60}
\eeq
where $E_{l}=Qz'$,
$D_{j/l}^{ind}$ is the induced radiation FF
(it depends on the parton energy $E$, but not virtuality), and 
$D_{l/i}^{DGLAP}$ is the vacuum DGLAP FF.

We have computed the DGLAP FFs with the help of the PYTHIA event 
generator \cite{PYTHIA}.
One gluon induced emission has been computed within the 
LCPI formalism \cite{LCPI} using 
the  method elaborated in \cite{Z04_RAA}. As in \cite{Z04_RAA,raa08}
we take $m_{q}=300$ and $m_{g}=400$ MeV for the quark and gluon quasiparticle masses.
Our method of calculation of the in-medium FF via the one gluon 
probability distribution is described in detail in \cite{raa08},
and need not to be repeated here. We just enumerate its basic
aspects. The multiple gluon emission
is accounted for employing Landau's method as 
in \cite{BDMS_RAA}.
For quarks the leakage of the probability 
to the unphysical region of $\Delta E> E$ is accounted for by  
renormalizing the FF.
We also take into account the $q\to g$ FF. Its normalization
is fixed from the momentum conservation for $q\to q$ and $q\to g$ 
transitions. The normalization of the $g\to g$ FF
is also fixed from the momentum sum rule.
The collisional energy loss, which is small \cite{Z_Ecoll}, is taken 
into account by renormalizing
the temperature of the QGP for the radiative FFs using the condition:
$\Delta E_{rad}(T^{\,'}_{0})=\Delta E_{rad}(T_{0})+\Delta E_{col}(T_{0})$, where
$\Delta E_{rad/col}$ is the radiative/collisional energy loss, $T_{0}$
is the real initial temperature of the QGP, and $T^{\,'}_{0}$ is the 
renormalized temperature.

We calculate the hard cross sections  using the LO 
pQCD formula.
To simulate the higher order $K$-factor
we take for the virtuality scale in $\alpha_{s}$ the value 
$cQ$ with $c=0.265$ as in the PYTHIA event generator \cite{PYTHIA}.
We account for the nuclear modification of the parton densities
(which leads to some small deviation of $R_{AA}$ from unity even without
parton energy loss) with the help of the 
EKS98 correction \cite{EKS98}.
For the vacuum FFs we use the KKP parametrization \cite{KKP}.

As in \cite{raa08}, we evaluate the induced gluon emission and 
the collisional energy loss for the running $\alpha_s$
frozen at some value $\alpha_{s}^{fr}$ at low momenta. For vacuum a
reasonable choice is $\alpha_{s}^{fr}\approx 0.7$.
This value was previously obtained 
by fitting the low-$x$ proton structure function $F_{2}$ within
the dipole BFKL equation \cite{NZ_HERA}. 
To study the role of the in-medium 
suppression of $\alpha_{s}$ we perform the computations for several
smaller values of $\alpha_{s}^{fr}$.

\vspace{.2cm}
\noindent{\bf 3.} 
We describe the QGP in the Bjorken model \cite{Bjorken2}
which gives $T_{0}^{3}\tau_{0}=T^{3}\tau$. We take $\tau_{0}=0.5$ fm.
To simplify numerical computations
for each impact parameter $b$ we 
neglect variation of the initial temperature $T_{0}$ in the 
transverse directions.
We evaluate its value using 
the entropy/multiplicity ratio
$dS/dy{\Big/}dN_{ch}/d\eta\approx 7.67$ obtained in \cite{BM-entropy}.
For the central $Au+Au$ collisions at $\sqrt{s}=200$ GeV
$T_{0}\approx 300$ MeV and for 
$Pb+Pb$ collisions at $\sqrt{s}=2.76$ TeV
$T_{0}\approx 400$ MeV. 
For the nuclear density we use the Woods-Saxon nucleus density
with parameters as in \cite{ALICE}.
The fast parton path length in the QGP, $L$, 
in the medium has been calculated according to the position
of the hard reaction in the impact parameter plane.
To take into account the fact that at times about $1-2$ units of 
$R_{A}$ the transverse expansion
should lead to fast cooling of the hot QCD matter \cite{Bjorken2} we also 
impose the condition $L< L_{max}$. We performed the computations for 
$L_{max}=8$ and 10 fm. The difference between these two versions is small. 

\vspace{.2cm}
\noindent{\bf 4.}
In Fig.~1 the theoretical $R_{AA}$ 
obtained for $\alpha_{s}^{fr}=0.7$, 0.6, and 0.5
for the chemically equilibrium 
and purely gluonic plasmas is compared to the PHENIX data \cite{PHENIX08} 
on $\pi^{0}$ production in the 0-5\% central $Au+Au$ collisions
at $\sqrt{s}=200$ GeV.
\begin{figure} [h]
\begin{center}
\epsfig{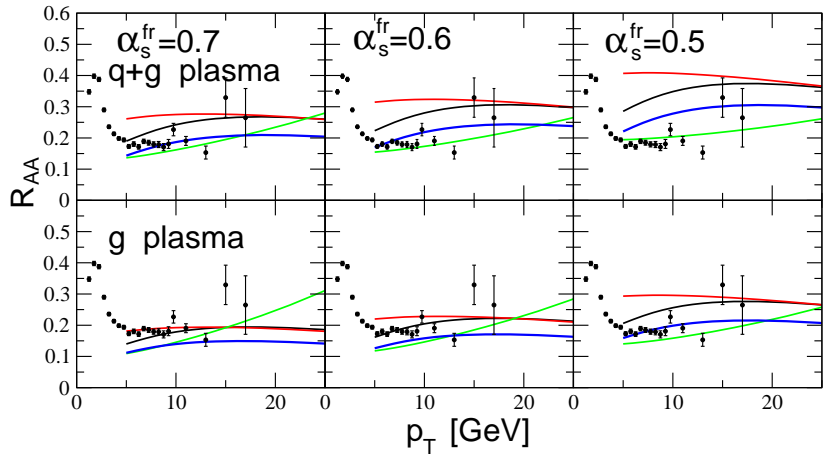}
\end{center}
\caption[.]
{
The factor $R_{AA}$ for $\pi^{0}$ production
in the 0-5\% central $Au+Au$ collisions
at $\sqrt{s}=200$ GeV for $\alpha_{s}^{fr}=0.7$, 0.6, and 0.5.
The upper panels are for the chemically equilibrium
plasma, and the lower ones for purely gluonic plasma.
Black line: the total radiative part (quarks plus gluons);
red line: the radiative quark part;
green line: the radiative gluon part;
blue line: the radiative (quarks and gluons) 
plus collisional, and plus 
energy gain due to gluon absorption.
The theoretical curves obtained for $L_{max}=8$ fm.
The experimental points are the PHENIX data \cite{PHENIX08}.
}
\end{figure}
The results are presented for radiative energy loss
and with inclusion of collisional energy loss and radiative energy
gain. The effect of the radiative energy gain on 
$R_{AA}$ is practically negligible  and can be safely neglected. 
\begin{figure} [h]
\begin{center}
\epsfig{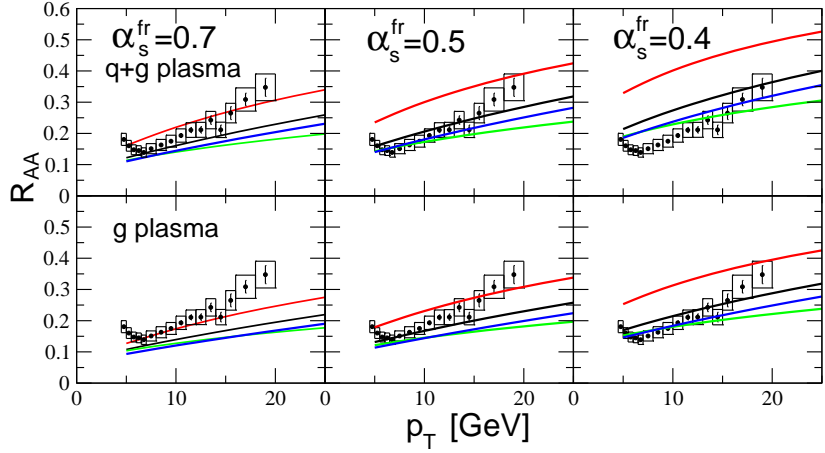}
\end{center}
\caption[.]
{
The same as in Fig.~1 for the charged hadrons in $Pb+Pb$ collisions
at $\sqrt{s}=2.76$ TeV for $\alpha_{s}^{fr}=0.7$, 0.5 and 0.4.
The experimental points are the ALICE data \cite{ALICE},
as in \cite{ALICE} the boxes contain the systematic errors.
}
\end{figure}
The growth of $R_{AA}$ for gluons in Fig.~1 is due to the $q\rightarrow g$ 
transition which is usually neglected. 
However, it does not affect strongly the total
$R_{AA}$ since for $\sqrt{s}=200$ GeV 
the gluon contribution to the hard cross section is small at $p_{T}\gsim
15$ GeV.
In Fig.~2 we compare our results for $\alpha_{s}^{fr}=0.7$, 0.5, and 0.4
with the ALICE data 
\cite{ALICE} for charged hadrons in $Pb+Pb$  collisions at $\sqrt{s}=2.76$ TeV. 

As can be seen from Figs.~1,~2, the collisional energy loss suppresses  
$R_{AA}$ only by about 15-25\%.
For the equilibrium plasma the data for $\sqrt{s}=200$ GeV can be 
described with $\alpha_{s}^{fr}\approx 0.6\div 0.7$. The data for 
$\sqrt{s}=2.76$ TeV agree better 
with $\alpha_{s}^{fr}\approx 0.4\div 0.5$.
It provides evidence for the thermal suppression of $\alpha_{s}$
at LHC due to higher temperature of the QGP.

\vspace{.2cm}
\noindent {\bf 5}. 
In summary, we have analyzed the 
data on $R_{AA}$ obtained in the PHENIX experiment on $Au+Au$ collisions
at $\sqrt{s}=200$ GeV \cite{PHENIX08} at RHIC and in the ALICE experiment on
$Pb+Pb$ collisions
at $\sqrt{s}=2.76$ TeV \cite{ALICE} at LHC.
Our results show that slow variation
of $R_{AA}$ from RHIC to LHC supports that the QCD coupling constant
becomes smaller in the hotter QGP at LHC. 

\section*{Acknowledgments}
I am grateful to the organizers for such an enjoyable and stimulating meeting
and for financial support of my participation.

\end{document}